RESEARCH ARTICLE

# Optimizing Scheduling Techniques for Enhanced Carrier Aggregation in LTE-Advanced Networks

Sajjad Emdadi Mahdimahalleh [1,*] and Vahid Tabataba Vakili [2]

## ABSTRACT

In this study, we investigate resource scheduling strategies for the downlink in LTE-Advanced networks, focusing specifically on systems that employ Carrier Aggregation (CA) with multiple Component Carriers (CCs). Effective scheduling is crucial in CA-enabled LTE-Advanced to maximize efficiency and support diverse user needs. We examine three conventional scheduling schemes—Joint User Scheduling (JUS), Separated Random User Scheduling (SRUS), and Separated Burst-Level Scheduling (SBLS). Although JUS yields superior performance, it is complex and does not factor in Quality of Experience (QoE) metrics. In contrast, SRUS and SBLS, while less computationally intensive, underutilize CA and fall short in resource allocation fairness. To overcome these limitations, we propose a new scheduling scheme, Quality of Service and Channel Scheduling (QSCS), which dynamically adjusts CC allocations at the burst level based on user service priority and signal quality. Simulation results show that QSCS achieves throughput similar to JUS while selecting optimal CCs based on Quality of Service (QoS) and channel parameters, resulting in enhanced QoE compared to other methods. This proposed approach demonstrates significant potential for improving resource allocation and user satisfaction in LTE-Advanced networks.

## 1. Introduction

One of the primary advancements in International Mobile Telecommunications (IMT), specifically in its IMT-Advanced iteration, is its capability to support peak data rates of 100 Mbps for high-mobility scenarios and up to 1 Gbps for low-mobility scenarios. As a progression of the third-generation (3G) Long-Term Evolution (LTE) system, LTE-Advanced aims to meet these heightened demands by operating over wider bandwidths than traditional 3G LTE systems [1]. A core feature of LTE-Advanced, known as Carrier Aggregation (CA), is integral to achieving these enhanced data rates outlined by IMT-Advanced standards [2]. CA in LTE-Advanced systems enables a single user to access up to 100 MHz of bandwidth while maintaining backward compatibility with earlier LTE versions. This backward compatibility is achieved by combining multiple LTE Component Carriers (CCs), each with a bandwidth of 20 MHz, to collectively provide up to 100 MHz for LTE-Advanced users [3].

In an LTE-Advanced system, both LTE-Advanced users and legacy LTE users can coexist, though they have different CC access capabilities [4]. Due to hardware limitations, LTE users are generally limited to accessing a single CC, while LTE-Advanced users can utilize multiple CCs [5]. Additionally, certain CCs may not support legacy LTE users, meaning that the enhanced NodeB (eNB) must identify the user type before assigning CCs. Following this initial user categorization, a suitable resource scheduling algorithm is employed to allocate CCs effectively.

Research on carrier scheduling (CS) schemes in LTE-Advanced systems with CA is extensive and diverse, often varying in system assumptions and models. These assumptions generally focus on two primary areas: the CA deployment scenario and the traffic model. Based on the CA deployment scenarios discussed in [6], three primary aggregation scenarios are common: intra-band contiguous CA, intra-band non-contiguous CA, and inter-band non-contiguous CA. Notably, in inter-band non-contiguous CA, CCs across different frequency bands may exhibit distinct radio propagation characteristics, making this scenario unique from intra-band CA and requiring specific considerations in CS studies. Existing literature, such as [7]–[9], provides insights into these scenarios. While high-frequency bands are typically preferred for LTE-Advanced [10], intra-band CA is often assumed due to the challenges of implementing inter-band CA in practical scenarios [11]–[15]. Traffic models, as recommended in [16], also play a role in performance evaluations of LTE-Advanced systems and generally follow two categories: time-variant and static user populations. Time-variant models, in which users arrive with finite transmission tasks and depart upon completion, are widely used, though [15] remains one of the few studies focusing specifically on CS performance within static user populations.

In JUS [1], the eNB calculates user throughput across all CCs, allowing for optimal resource distribution. However, as user and CC counts increase, JUS's high computational complexity poses a challenge, and it lacks considerations related to the QoE for users. Alternatively, SRUS and SBLS [1], [15] are less complex and focus on calculating throughput in a single CC, although they, too, present limitations in CA support and resource fairness.

Given these factors, an effective carrier scheduling scheme capable of managing multiple CCs and optimizing user allocation is crucial for LTE-Advanced with CA. Furthermore, with the rising importance of QoE—defined as service quality as perceived subjectively by the user—network operators face mounting pressure to deliver consistently high-quality services across heterogeneous networks. Users now expect seamless, high-quality service at any time, location, and on any device, creating a convergence need from the user's perspective.

From a network perspective, user-perceived quality is impacted by various end-to-end service delivery factors, including network performance, hardware capabilities, data encoding, and protocols. Each end-user requirement translates into specific technical and business challenges that shape the provisioning of end-to-end services. As a result, modern services must operate across different media, networks, and service providers to meet these demands [17].

In this work, we introduce a novel scheduling algorithm, termed QSCS, with the objective of enhancing QoE in LTE-Advanced systems. Numerous parameters influence QoE, and the impact of these parameters is often context-dependent. Simulation results indicate that QSCS improves both QoS and Quality of Channel (QoC). Prior to presenting the system model, we outline key parameters used in the proposed algorithm, followed by a detailed discussion of simulation results and findings.

## 2. QoS and QoC Considerations

QoS is defined by the Internet Engineering Task Force (IETF) as "a set of service requirements to be met by the network while transporting a flow" [18]. In this study, QCI parameters have been considered as core service requirements, and achieving these standards is one of the primary targets [19]. Simulation results indicate that the proposed QSCS algorithm shows significant improvements in QoS compared to other scheduling schemes.

QoC, as considered in this study, refers to the signal-to-noise ratio calculated based on user distance from the cell and the allocated CC. As an additional parameter aimed at enhancing user satisfaction, QoC plays a crucial role in the new scheduling scheme. By selecting the optimal CC based on user proximity, QSCS demonstrates considerable improvements in channel quality. Simulation results confirm that users experience stronger signal strength and improved channel quality with QSCS over other methods.

Thus, users can access services with higher quality and improved signal strength, resulting in a better QoE with QSCS than with other algorithms used for comparison.

## 3. QoS Class Identifier

The QoS Class Identifier (QCI) is a scalar value that serves as a reference for specific packet forwarding behaviors, such as packet loss rate and delay budget. Within the access network, QCI parameters are mapped to node-specific controls that manage packet forwarding behavior, including scheduling weights, admission, and queue management thresholds, and link layer configurations, which are typically pre-set by the network operator at designated nodes (e.g., eNodeB) [19].

QCI is a key QoS parameter utilized in QSCS, guiding the allocation of CCs to each user based on their service requirements. Each Service Data Flow (SDF) is uniquely associated with one QCI. Table I [19] provides standardized characteristics for various QCI levels, including their priorities and permissible delay values. The QSCS algorithm leverages this table to allocate resources efficiently, ensuring that users receive services aligned with their specific requests.

Services utilizing a Guaranteed Bit Rate (GBR) QCI and operating at or below the GBR threshold can generally expect minimal packet loss due to congestion. Under these conditions, 98% of packets should experience delays within the QCI's specified Packet Delay Budget (PDB). However, if delays exceed the PDB, the QSCS mechanism may not meet the required service level. In contrast, services using a Non-GBR QCI should anticipate potential packet drops caused by congestion. For packets that are not dropped, 98% should experience delays within the QCI's PDB despite congestion-related challenges.

## 4. System Model

We consider a downlink CA system based on Orthogonal Frequency-Division Multiplexing (OFDM), consisting of a single Node B (eNB) and multiple User Equipment (UEs), as illustrated in Fig. 1 [15]. The UEs are distributed randomly and uniformly within the cell.

In this setup, we assume a non-contiguous CA scenario (inter-band non-contiguous CA) with $M$ CCs positioned in different frequency bands, each potentially with distinct bandwidths. The system comprises two types of users: LTE-A users and legacy LTE users. To streamline the analysis, we categorize CCs into two main types. LTE-A-specific CCs are exclusively accessible to LTE-A users, while shared CCs are available to both LTE-A and LTE users. Fig. 2 shows the structure of the QSCS resource scheduler framework, highlighting the allocation approach based on user type and channel conditions.

TABLE I: Standardized QCI Characteristics (adapted from [18])

| QCI | Resource type | Priority | Packet delay budget | Example services |
|---|---|---|---|---|
| 1 | GBR | 2 | 100 ms | Conversational voice |
| 2 |  | 4 | 150 ms | Conversational video (live streaming) |
| 3 |  | 3 | 50 ms | Real time gaming |
| 4 |  | 5 | 300 ms | Non-conversational video (buffered streaming) |
| 5 |  | 1 | 100 ms | IMS Signalling |
| 6 |  | 6 | 300 ms | Video (buffered streaming), TCP-based (e.g., www, e-mail, chat, ftp, p2p file sharing, progressive video, etc.) |
| 7 | Non-GBR | 7 | 100 ms | Voice, video (live streaming), interactive gaming |
| 8 |  | 8 | 300 ms | Video (buffered streaming), TCP-based (e.g., www, e-mail, chat, ftp, p2p file sharing, progressive video, etc.) |
| 9 |  | 9 |  |  |

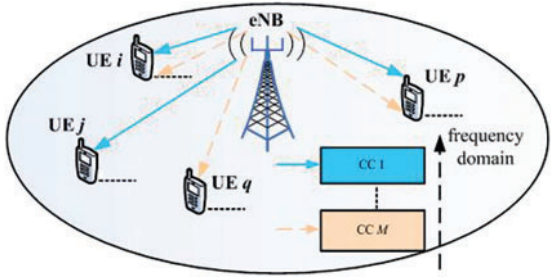

Fig. 1. The downlink CA system with M CCs.

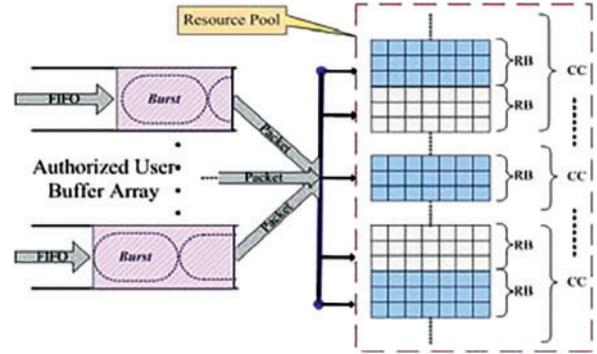

Fig. 3. Illustration of JUS resource scheduler.

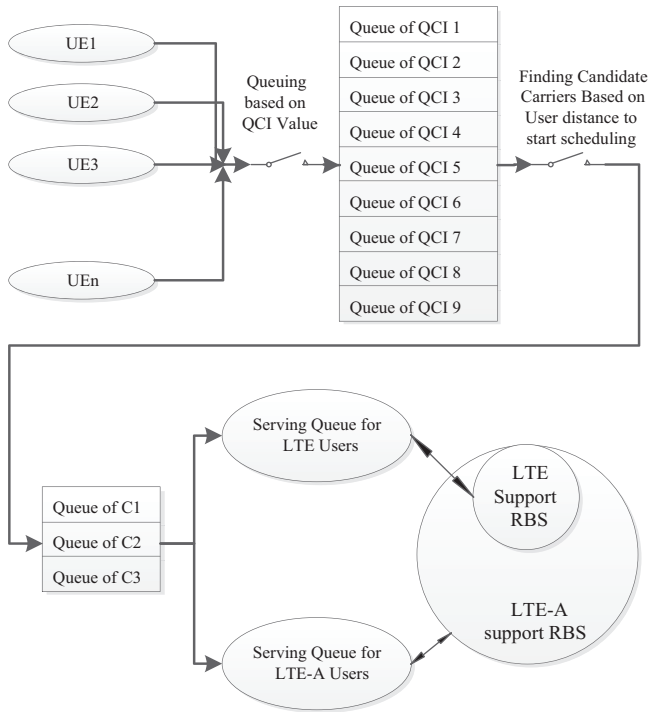

Fig. 2. The framework of the QSCS resource scheduler.

When a user arrives, the system determines which users to serve based on the requested QCI value and allocates optimized CCs according to the user's distance from the cell. For LTE users, the system selects a single CC, with bandwidth ranging from 1.4 MHz to 20 MHz. In contrast, for LTE-A users, all available carriers can be assigned, provided that the total number of aggregated CCs does not exceed five.

Once users are assigned to specific CCs, resource scheduling occurs within each carrier. During scheduling, each CC comprises $M$ physical resource blocks (PRBs) allocated to users, with the PRB serving as the smallest allocation unit. Each PRB corresponds to a single time slot in the time domain and occupies 180 kHz in the frequency domain [20]. The number of PRBs per CC is bandwidth-dependent, as detailed in Table II [21], [22].

At the eNB, traffic for each UE is buffered in separate first-in, first-out (FIFO) queues, organized by QCI, and following various multi-user scheduling schemes. This study focuses on elastic data streams. At the start of each frame, the Resource Scheduler (RS) assigns Resource Blocks (RBs) to authorized UEs in the queue.

## 5. Multi-user Scheduling Scheme

Building on the previously outlined system model, this section provides a concise overview of five distinct multi-user scheduling schemes. For each scheme, we discuss its operational procedures along with an analysis of its advantages and limitations. In this context, "UE" refers interchangeably to the user.

### 5.1. JUS Algorithm

In the JS framework, the RBs across all CCs within the system are combined into a unified resource pool. This aggregated pool is managed by a single RS, as illustrated in Fig. 3 [15].

TABLE II: TRANSMISSION BANDWIDTH CONFIGURATION

| Channel BW (MHz) | 1.4 | 3 | 5 | 10 | 15 | 20 |
|---|---|---|---|---|---|---|
| No. of PRB | 6 | 15 | 25 | 50 | 75 | 100 |

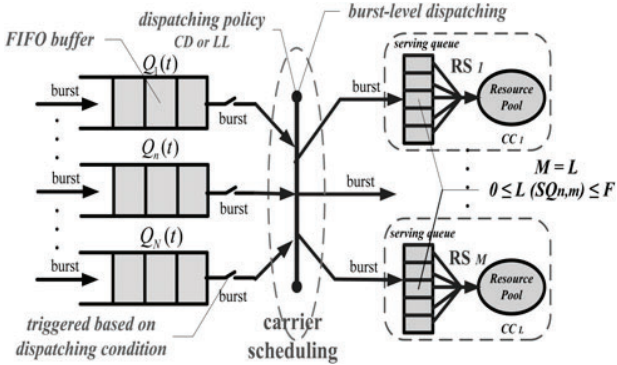

Fig. 4. Illustration of SBLS resource scheduler.

On the other hand, there is only a one-level scheduler in JS, meaning no specific user allocation method is applied. This implies that users in the system may receive data transmitted from all available CCs simultaneously, which may not be practical. Consequently, the feasibility of JS is largely impacted by increased user complexity and the absence of policies to ensure good QoS and QoC. Nevertheless, it remains an optimal approach for achieving maximum spectral efficiency and full system resource utilization.

### 5.2. SRUS Algorithm

The SRUS scheme operates with a two-level scheduling structure. The first level is responsible for assigning each user to only one of the available CCs, utilizing an instinctively designed random dispatcher. This structure aims to ensure load balancing, where each CC is required to handle a fair share of user transmissions, as per [15]. At the second level, an independent RS is allocated to each CC, as referenced in Section 1. Within the resource pool, all RBs assigned to an individual user are confined to a single CC, effectively reducing user-side complexity by eliminating the need for broader bandwidth support.

However, a potential limitation arises in scenarios where certain CCs remain idle once their designated workload is completed, even if other CCs are still active. This can lead to reduced trunking efficiency within the SRUS scheme. Moreover, there is an absence of specific QoS or QoC policies, which may impact the service quality provided to users.

### 5.3. SBLS Algorithm

The SBLS framework, as illustrated in Fig. 4, treats each CC as an independent carrier, similar to SRUS. Accordingly, the number of RSs and the size of each RS's resource pool are consistent with SRUS, and it also employs a two-level scheduling structure. The key distinction in SBLS lies in user allocation; users are not served by a fixed RS, allowing the dominant RS for a user to change at the burst level, though only one RS is active at a time.

In SBLS, only the burst at the front of the buffer is dispatched, with the dispatching governed by specific policies. Two dispatching policies are applied here: circular dispatching (CD) and least load (LL). Under CD, the dominant RS is selected cyclically for each burst, regardless of the user's affiliation. Conversely, with LL, the RS is chosen based on the minimum load of the CC relative to its BW.

The SBLS approach enhances resource utilization by decreasing the granularity of traffic dispatching, resulting in better traffic load distribution across CCs. While this finer dispatching granularity improves resource use compared to SRUS, it operates at the burst level, which is less granular than SRUS's user-level allocation. Therefore, SBLS offers higher resource utilization over SRUS, although it cannot outperform JUS. As with previous algorithms, no QoS or QoC policy is in place to manage service quality for users.

From a complexity standpoint, the SBLS algorithm is more intricate than SRUS due to its burst-level dispatching mechanism and dynamic RS allocation. However, when compared to JUS, SBLS remains relatively simple. In terms of fairness, SBLS offers a more balanced resource allocation than SRUS, making it a more equitable option in multi-user scenarios. Nonetheless, both SBLS and SRUS lack mechanisms to address quality parameters directly affecting end users, such as QoS and QoC, which may limit their effectiveness in environments with stringent quality requirements.

### 5.4. QSCS Algorithm

The proposed QSCS Scheduling algorithm, similar to SBLS and SRUS, follows a two-level scheduling structure. QSCS takes different carrier's frequency bands and, based on them, defines different coverage areas, and all PRBs of one area would be aggregated in one pool. Fig. 5 illustrates different coverage areas of one sample network.

Also, QSCS takes the user's requests QCI value, and then different queues according to each QCI value would be defined. The other parameter that QSCS checks is the user's distance to the cell, and with this location status, CCs will be prioritized to be dedicated to users.

The aggregation scenario that QSCS uses is inter-band non-contiguous CA, and its target is allocating optimized CCs for providing services with the highest signal-to-noise ratio according to the user's location and also in accordance with considerations of QoS that have been mentioned in Table I. QSCS allows users to use aggregated CCs, and the maximum number of CCs that are allowed to be aggregated in this algorithm is five. It means that if there are 5 carriers in the network with 20 MHz BW, by deploying this new RS algorithm, 100 MHz BW can be assigned to one user, and based on LTE-A definition, reaching 1 Gbps speed of data for a fixed user would be feasible.

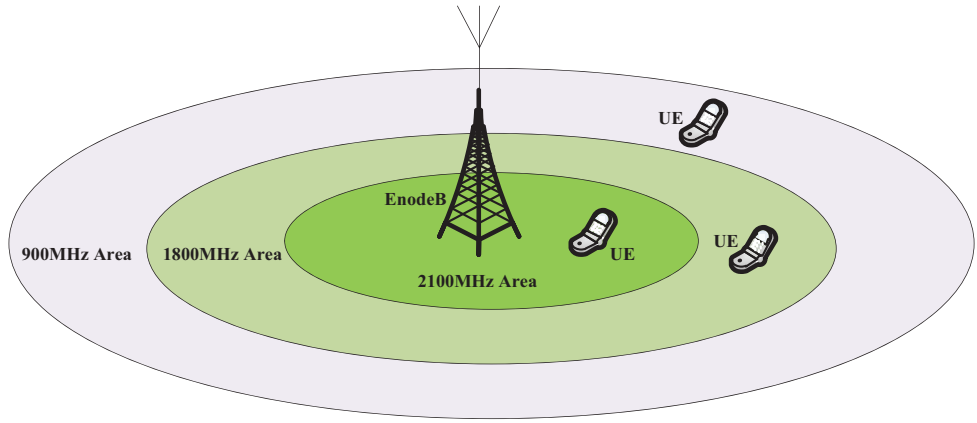

Fig. 5. Illustration of different coverage areas.

In the QSCS algorithm, users, in accordance with their request's priority and their arrival time to the queue (first-in, first-out), take resource blocks. All considerations related to GBR and non-GBR services (Table I) are included in the new proposed algorithm.

For example, if a user sends a request with $2^{nd}$ priority and there is another user whose service's priority is 7, and all PRBs are busy, the system will interrupt that service with $7^{th}$ priority (lower priority non-GBR service) and give its resource blocks to this new user's request with higher priority.

Also, if a new user arrives and no optimized CCs based QoC's considerations exist, system in each scheduling time checks to find optimized free CCs and change the current CC to optimized CC.

According to the definition of QSCS, it is expected to see a considerable improvement in QoS and QoC that users will experience, and users should have more satisfaction by deploying this new scheduling algorithm.

## 6. Simulation Results and Analysis

This section presents downlink system-level simulations conducted to validate the analysis discussed in Section 2. Prior to presenting the results, an overview of the simulation assumptions and definitions of performance metrics is provided. Some simulation parameters are derived from the 3G LTE specifications [20] and are summarized in Table III.

### 6.1. Simulation Assumptions
1. *CC configuration*: The configuration includes a total of six CCs across three frequency bands. Two CCs operate in the 900 MHz band with bandwidths of 1.4 MHz and 5 MHz, respectively. In the 1800 MHz band, there are three CCs, two with a bandwidth of 1.4 MHz and one with 3 MHz. The sixth CC is positioned in the 2100 MHz band with a bandwidth of 1.4 MHz.
2. *Users Type*: It has been assumed that all users are LTE-A users that can take CCs from different frequencies and work with them simultaneously.

TABLE III: Simulation Parameters

| | |
|---|---|
| User density per sector | 10 |
| Minimum separation between UE and eNB | 35 m |
| Frame duration | 1 ms |
| Distance between subcarriers | 15 KHz |
| Simulation time | 3 min |
| Type of users | LTE-A |
| QCI values | 1–9 |
| Max. time of a requested service | 2500 ms |
| No. of existing CCs | 6 |
| No. of frequency bands | 3 |
| Avg. data rate of each request | 35 mbps |
| Avg. no. of requests per user | 25 |

3. *Traffic Type*: It has been considered elastic and also one user can have multiple requests. Maximum number of requests per user assumed to be 25.
4. *Services time*: Users can ask for services with different periods, and the maximum time of any serice has considered to be 2500 ms.
5. *User's location*: When a user arrive to cell coverage area, according to its location, proper list of CCs will be provided to allocate to user in case of service's request.

### 6.2. Performance Metrics
The primary objective of the simulations is to assess the performance of different carrier scheduling (CS) schemes with respect to resource utilization. This evaluation is conducted based on the following three metrics:

1. *Average Sojourn Time:* This metric represents the time interval between the arrival and full departure of a data burst. The sojourn time is calculated by recording the allocation time of the first resource block (RB) and the allocation time of the last RB for the data burst. The duration between these two timestamps defines the sojourn time.
2. *Best Signal Allocation:* When a user moves through different coverage areas within a cell, the optimal component carrier (CC) for that user is determined based on which CC provides the highest signal-to-noise ratio (SNR). This metric compares all physical resource blocks (PRBs) assigned by different algorithms to ensure that the most suitable CCs are allocated.

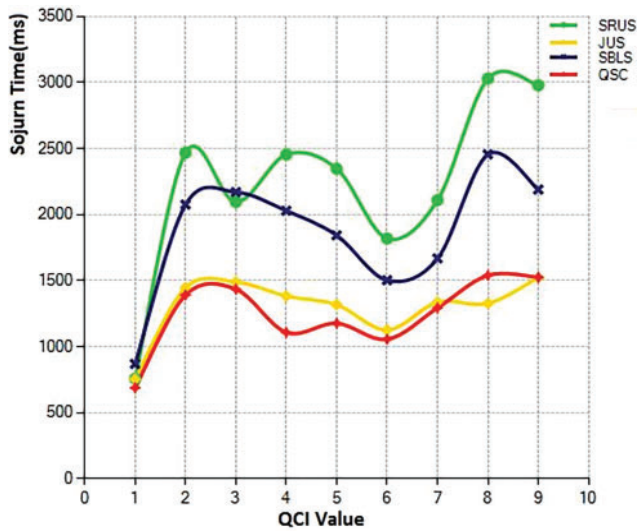

Fig. 6. Sojourn time vs. QCI values of services in different CS schemes.

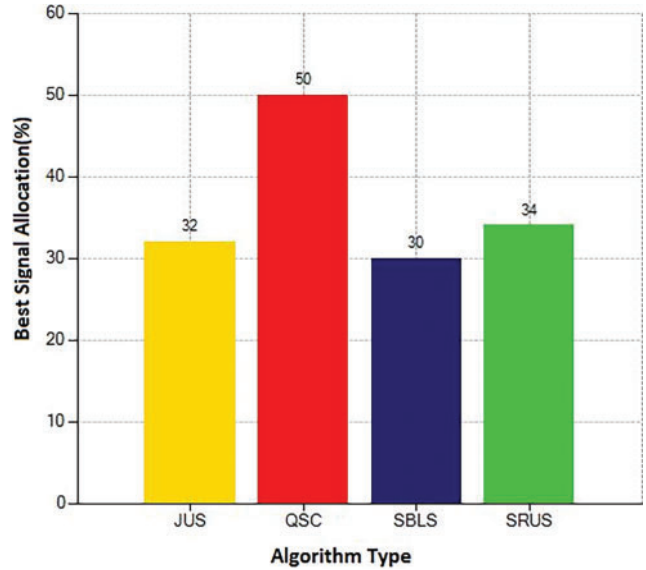

Fig. 7. Best allocated CCs vs. CS algorithms type.

3. *Quality of Guaranteed Services:* According to the QCI table, services with QCI values between two and five are guaranteed bit rate (GBR) services. When a user requests a service with these priority levels, the system must allocate the required RBs and ensure they are available for the duration of the service.

### 6.3. Performance Comparison Between QSCS and the Reference Schemes

As analyzed in Section 2, under conditions of elastic traffic input, the objective of QSCS is to enhance the quality of experience, which translates to higher QoS under equivalent conditions compared to other algorithms, along with optimized signal strength. Since QSCS, similar to JUS, utilizes carrier aggregation (CA), it is expected to surpass JUS in performance. The simulation results illustrated in Fig. 6 depict the sojourn time for the evaluated CS schemes across varying QCI values.

From Fig. 6, it is evident that QSCS achieves a significant performance gain over other CS schemes, particularly when compared to SRUS and SBLS and for example in QCI = 4, is at least 110% better than SRUS and 90% better than SBLS and also 20% better than JUS. There is just one point in QCI = 8 that JUS is performing better than QSCS. As already discussed previously in the same conditions, when there are requests by priority between two and five and this time, also other requests by lower priority exist, the system will assign resources to the services that have the highest priority, and they are GBR services, even if it couldn't find any free resources it interrupts other services with lower priority and gives their resources to these request. In this simulation, the same issue has happened, and resources have been assigned to the higher priority requests.

These results align with expectations. The differences in resource utilization between QSCS and the three other reference CS schemes are illustrated more explicitly in Fig. 7, showcasing the optimization of allocated CCs based on user locations within coverage areas. From Fig. 7, we can see that QSCS has the highest optimized CCs that have been assigned and is at least 16% better than other schemes. The reason that it doesn't reach 100% is coming from resource limitations. If the system had enough resources, it would have higher performance.

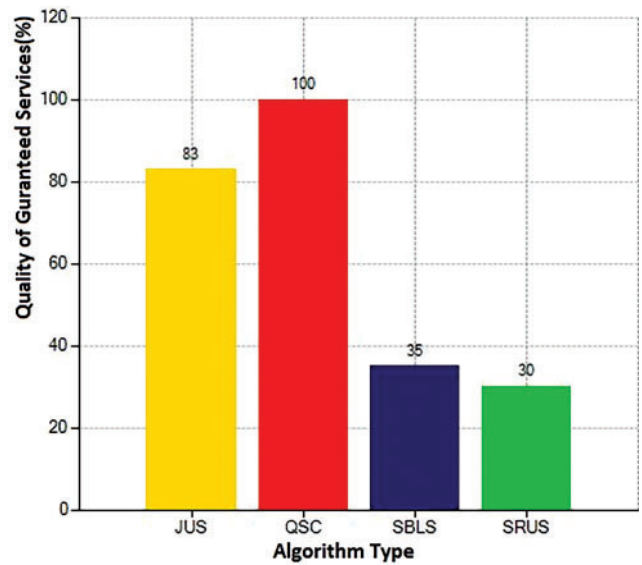

Fig. 8. QoS of services with QCI values 2–5 in different CS Schemes.

Also, from Quality point of view, the most comparable part of new scheduling algorithm is separation of requests based on their relevant QCI values. QSCS checks the type of services and accordingly provides the required resources for users. As Fig. 8.

Illustrates for utilizing existing resources, as QSCS provides guaranteed bit rates from start time till end time for these requests, GBR services that users have requested, are compared in this figure. It's clear that QSCS should give 100% of the required BW, but as the other scheduling schemes don't care to this point, their quality of service for requests with QCI value of two to five can't be as much as in QSCS.

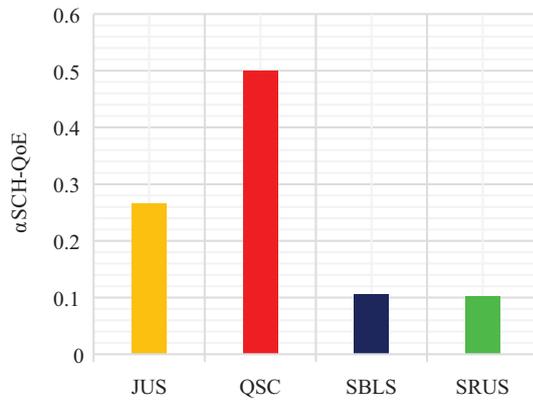

Fig 9. QoS of services with QCI values 2–5 in different CS Schemes.

## 7. Conclusion

This paper proposes and analyzes a novel carrier scheduling scheme, namely the QSCS scheme. Under the assumption of inter-band carrier aggregation, the proposed scheduling algorithm achieves a favorable trade-off between the three reference CS algorithms in terms of both quality of experience and system complexity. Simulation results demonstrate a 24% improvement in QoE. However, this improvement is accompanied by an increase in system complexity attributed to the deployment of CA and additional control conditions.

For comparing QoE parameters that users experience in the network with different scheduling algorithms one coefficient has been defined as $\alpha_{\text{SCH-QoE}}$ that is termed as scheduling QoE coefficient that will be calculated by multiplying of QoS coefficient ($\beta_{\text{QoS}}$) and QoC coefficient ($\beta_{\text{Opt.CC}}$). These two last parameters are being calculated by the results of Figs. 7 and 8:

$$\alpha_{\text{SCH-QoE}} = \beta_{\text{QoS}} \times \beta_{\text{Opt.CC}}$$

According to this new parameter definition Fig. 9 illustrates QoE comparison of different scheduling schemes.

As Fig. 9 proves, QSCS has the highest QoE coefficient, and its performance is 24% better than other scheduling algorithms, this factor, as the major factor, is very important for mobile service providers because it actually shows a percentage of customer satisfaction.

## Conflicts of Interest

The authors declare that they do not have any conflict of interest.